# Phase Change Memory by GeSbTe Electrodeposition in Crossbar Arrays

Yasir J. Noori[1,*], Lingcong Meng[2], Ayoub H. Jaafar[1], Wenjian Zhang[2], Gabriela P. Kissling[2], Yisong Han[3], Nema Abdelazim[1], Mehrdad Alibouri[4], Kathleen LeBlanc[4], Nikolay Zhelev[2], Ruomeng Huang[1], Richard Beanland[3], David C. Smith[4], Gillian Reid[2], Kees de Groot[1,*], Philip N. Bartlett[2,*].

[1]School of Electronics and Computer Science, University of Southampton, Southampton, SO17 1BJ, UK
[2]School of Chemistry, University of Southampton, Southampton, SO17 1BJ, UK
[3]Department of Physics, University of Warwick, Coventry, CV4 7AL, UK
[4]School of Physics, University of Southampton, Southampton, SO17 1BJ, UK
*y.j.noori@soton.ac.uk; chdg@soton.ac.uk; p.n.bartlett@soton.ac.uk

**Abstract**

Phase change memory is an emerging type of non-volatile memory that shows a strong presence in the data-storage market. This technology has also recently attracted significant research interest in the development of non-Von Neumann computing architectures such as in-memory and neuromorphic computing. Research in these areas has been primarily motivated by the scalability potential of phase change materials in crossbar architectures and their compatibility with industrial nanofabrication processes. In this work, we have developed crossbar phase change memory arrays through the electrodeposition of GeSbTe (GST). We show that GST can be electrodeposited in nanofabricated TiN crossbar arrays using a scalable process. Various characterisation techniques, such as AFM, TEM and EDX were used to study the electrodeposited materials in these arrays. Phase switching tests of the electrodeposited materials have shown a resistance switching ratio of 2 orders of magnitude with an endurance of around 80 cycles. Demonstrating crossbar phase change memories via electrodeposition paves the way towards using this technique for developing scalable memory arrays involving electrodeposited materials for passive selectors and phase switching devices.

## 1. Introduction

Since phase change materials (PCMs) were first discovered by Ovshinsky in the 1960s, their most important application has been in optical data storage discs.[1–3] During the past decade, applications of PCMs have expanded to electronic memory devices. These are anticipated to bridge the gap between the low-density Dynamics Random Access Memories (DRAMs) and the slow Flash technology in traditional Von Neumann computing architectures.[4–6] Furthermore, the unique optical and electronic properties of PCMs, especially semiconductor metal chalcogenides, have recently been exploited in non-Von Neumann architectures such as neuromorphic and in-memory computing.[7,8] These works have shown that there is an important role that awaits PCMs in the future development of novel electronic as well as photonic technologies.[9,10] However, significant scaling in PCM production must be achieved to meet the high density demand for today's data storage industries or artificial neural networks. Traditional "mushroom" shaped phase change memory arrays suffer from thermal cross-talk between adjacent devices, thus confining PCM devices in nano cells is more favourable for high density scaling as it could reduce the thermal cross-talk and reduce the current required for switching operations.[11,12]

Plasma sputtering has commonly been used for depositing PCMs demonstrating strong switching properties.[13–17] Atomic layer deposition (ALD) has shown good composition control over the PCM in high aspect ratio structures.[18–21] Other PCM deposition methods that have been explored include chemical vapour deposition.[22,23] However, there are several limitations associated with these methods. For example, sputtering deposits material at an angle making it unsuitable in filling high density or high-aspect ratio structures. This limits its potential in scaling the density of memory arrays. ALD has low throughput due to very slow deposition rate and provides little spatial control over the deposition location. Furthermore, sputtering and CVD normally introduce surface and structural defects on pre-existing materials on the substrate due to ion bombardment or high temperature growth. This can cause severe damage to delicate ultrathin and 2D materials.[24–26] The deposition of PCMs over 2D materials has been used to reduce the switching power in phase change memory devices.[27,28]

We have recently shown that the electrochemical deposition (electrodeposition) method can produce large-area thin-films of GeSbTe (GST).[29] We demonstrated that the elemental composition of GST films can be altered via modulating the precursor concentration of the electrodeposition solution. We have also previously shown that our PCM thin films exhibit reasonable switching endurance, making them suitable for the future development of phase change memory technology.[30–32] Electrodeposition is a well-established material production method in the electronics industry. Although the electrodeposition of functional chalcogenides is far less mature than physical vapour deposition, electrodeposition has several unique advantages to offer. It is a bottom-up deposition technique that can selectively produce nanoscale materials over conductive patterns. Unlike sputtering, electrodeposition is not a line-of-sight deposition method as it can be used to deposit materials over 3D surfaces such as integrated photonic waveguides.[33,34] The ability to electrodeposit in lateral etched back cavities would open up a much larger design space for such memory than can be achieved with line-of-sight techniques. Secondly, it does not require inserting the target substrate in a harsh environment such as plasma or high temperatures. In previous work, we have shown that MoS$_2$ 2D materials can be electrodeposited over graphene electrodes while maintaining the latter's original quality.[35] Thirdly, it is capable of depositing materials within fabricated structures of high density and high aspect ratios as has long been practiced in the semiconductor industry using the Damascene process.[36] Unlike the latter process, which requires a uniform seed layer to act as an electrode and is mainly used for creating interconnects, our process uses individually accessible and addressable patterned lines to deposit functional phase switching materials. This

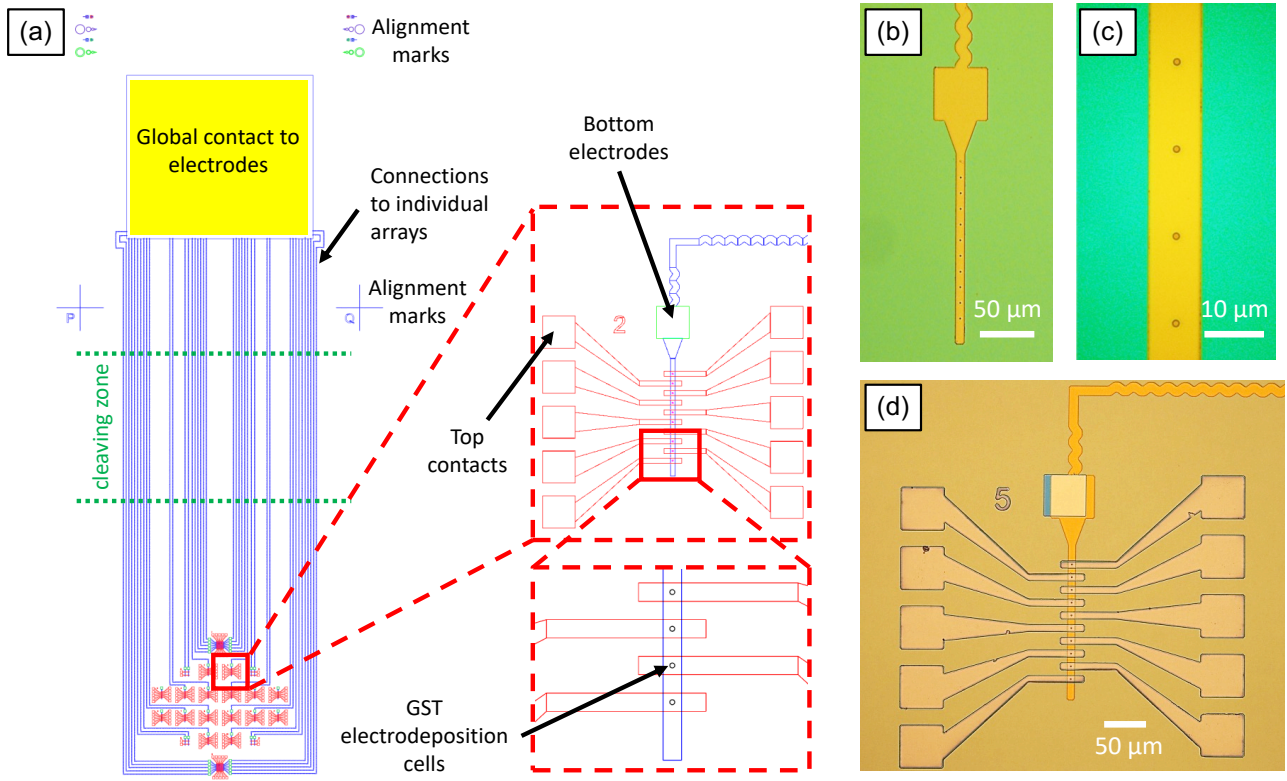

Figure 1 (a) A schematic of the mask layout of the crossbar chip design. The bottom memory electrodes are connected through long zigzag lines to the global contact where the electrodeposition potentiostat is clipped. The insets show schematic of the crossbar arrays. (b) and (c) Optical microscopy images of individual cells after fabrication. (d) An image of a finished crossbar array after material deposition and top contact fabrication.

property of electrodeposition can provide an important advantage in scaling the production of phase change memory devices into very dense arrays of sub 100 nm cells. This also enables the realisation of true 3D phase change memory architectures where the number of lithographic steps does not scale with the number of layers.

Electrodeposition has been exploited for depositing functional materials in crossbar arrays for resistive switching and thermoelectric devices, but neither process is scalable to submicron dimensions.[37,38] In this work, we report the development of nanofabricated phase change memories in crossbar arrays via the electrodeposition of ternary PCMs. Characterization of the deposits was performed using various microscopic and spectroscopic techniques to study the electrodeposition of GST in confined cells and quantify the composition of the produced deposits. Transmission Electron Microscopy (TEM) was used to image our fabricated phase change memories and characterize the materials as-deposited and after phase cycling. Finally, results from phase cycling the memory devices are presented, demonstrating the potential of electrodeposition in realizing phase change memory systems.

## Experimental Section

### Device fabrication

A confined cell geometry was fabricated in which the electrodeposited material is sandwiched between TiN electrodes. Figure 1 (a) shows a schematic of the mask layout that was used in fabricating the PCM devices. Details of each process step are displayed in supplementary Figure S1. The bottom TiN electrodes of the arrays, outlined in blue in Figure 1 (a), are designed such that they are all connected to a large pad which is Cr/Au coated before electrodeposition (highlighted in yellow). This pad acts as a global contact through which a potential can be applied to all the bottom electrodes on the chip to allow the PCM to be electrodeposited inside the memory cells. The cells, outlined in black, were fabricated using e-beam lithography and act as the "working electrode" during the electrodeposition process. Details of the fabrication processes can be found in the caption of Supporting Figure S1. Figure 1 (b) and (c) show microscope images of the fabricated TiN bottom electrodes and the memory cells prior to electrodeposition. Figure S2 shows higher magnification SEM images of the fabricated cells. After electrodeposition, the top electrodes, outlined in red in Figure 1 (a), were fabricated using photolithography and TiN lift-off processes, in a similar approach to fabricating the bottom electrodes. Finally, a photolithographic step was used to etch away the $SiO_2$ that covers the bottom electrode. The pattern of the latter step is outlined in green in Figure 1 (a). Figure 1 (d) shows an image of a fully fabricated crossbar memory array after material electrodeposition. A higher magnification image of Figure 1 (d) is presented in Figure S3.

### GeSbTe electrodeposition

All of the electrodeposition experiments were performed inside a controlled environment glovebox equipped with a nitrogen circulation system to ensure minimum water (<18 ppm) and oxygen levels (<5 ppm). Figure 2 (a) shows an illustrative example of the electrodeposition setup involving a potentiostat connected to the TiN electrodes through the Cr/Au global contact. The depositions were performed in a three-electrode cell that uses a Pt : Ir (90% : 10%) disc as a counter electrode and a Ag/AgCl wire in a glass frit (containing 0.1 M [N$^n$Bu$_4$]Cl in dichloromethane ($CH_2Cl_2$) as a reference electrode. Supplementary Figure S4 (a) shows an image of the electrodeposition setup. An electrolyte system has been synthesised in-house based on tetrabutylammonium

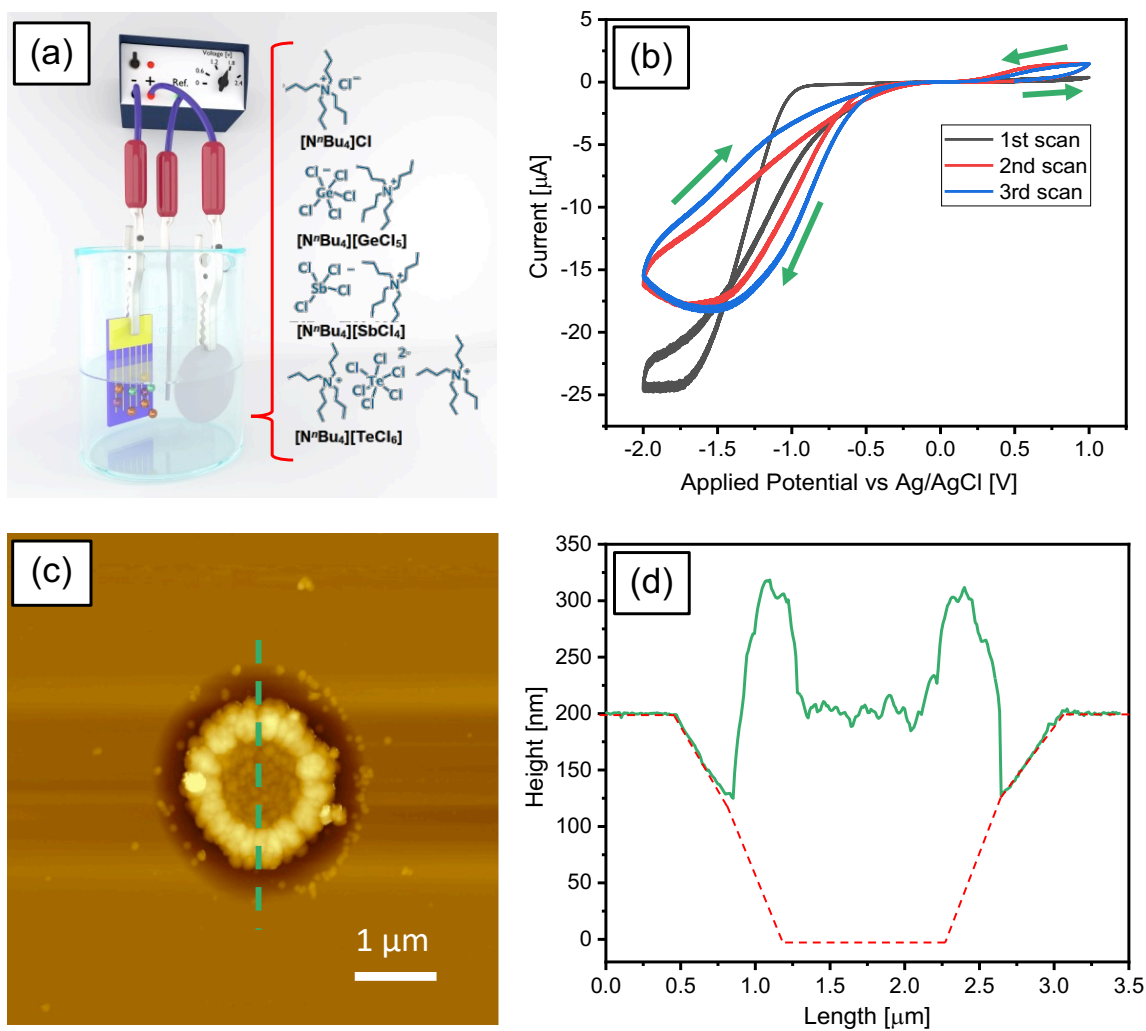

Figure 2 (a) An illustration diagram of a three-electrode electrochemical cell containing the crossbar array chip connected through its global electrode to the potentiostat; together with molecular structure of the precursors used in the dichloromethane solvent. (b) The consecutive CV scans taken using a TiN microelectrode array in a solution containing 2.5 mM of [N$^n$Bu$_4$] GeCl$_6$, 1 mM [N$^n$Bu$_4$] SbCl$_4$ and 2 mM [N$^n$Bu$_4$]$_2$TeCl$_6$. (c) An AFM image of a deposit inside a cell. (d) An AFM topography line profile along the deposit. The dashed line represents roughly the profile of an unfilled cell.

chlorometallate salts that can dissolve and dissociate readily in the weakly coordinating CH$_2$Cl$_2$ solvent. These well-defined and readily synthesizable reagents show very high solubility in organic solvents, which allows preparing electrolytes with a range of precursor concentrations. CH$_2$Cl$_2$ has several advantages over aqueous solvents. First, it is stable over a wider electrochemical window than water, making it suitable for electrodepositing materials that reduce at relatively high negative potentials, such as Ge. Second, it has low surface tension, which makes it a promising solvent for electrodeposition inside nano-patterned structures. In this work, the salts used to deposit Ge, Sb and Te are 2.5 mM [N$^n$Bu$_4$][GeCl$_5$], 1mM [N$^n$Bu$_4$][SbCl$_4$] and 2mM [N$^n$Bu$_4$]$_2$[TeCl$_6$], respectively. These precursor complexes are chemically very similar salts and mutually compatible in solution, allowing alloys of these materials to be deposited by varying the relevant chlorometallate concentrations. 0.1 M [N$^n$Bu$_4$]Cl was used as the supporting electrolyte. We have previously shown that electrodeposition solutions prepared from tetrabutylammonium chlorometallate, [N$^n$Bu$_4$]$_x$[MCl$_z$], electrolytes combined with tetrabutylammonium chloride, [N$^n$Bu$_4$]Cl as the supporting electrolyte, can be used to electrodeposit a wide variety of p-block elements, including selenium, indium, antimony, tellurium, germanium and bismuth.[39,40] This approach can be further expanded to include more p-block metals if their corresponding halometallate salts can be synthesised.

To understand the electrochemical deposition behaviour of this ternary metal electrolyte, cyclic voltammetry (CV) scans were carried out using TiN macro and micro electrodes. Supplementary Figure S4 (b) presents CV scans obtained from a disc shaped TiN electrode of 4 mm diameter which was defined on a chip by photolithography. The scans show clear changes in the current at cathodic potentials. These correspond to the reduction of Ge$^{IV}$, Sb$^{III}$ and Te$^{IV}$ to Ge$^0$, Sb$^0$ and Te$^0$. Further details can be found in the Supplementary Information. This is supported by our previous report which also included CV scans from individual components of the electrolyte.[29] Figure 2 (b) presents CV scans taken from a large array of 1 µm diameter TiN cells. In the first CV scan, a reductive limiting current of -25 µA at -1.7 V vs Ag/AgCl is observed. In the reverse scan, a current crossover at -1.3 V with no obvious stripping current is observed. The 2$^{nd}$ and 3$^{rd}$ cycles have relatively smaller currents compared to those observed from the 1$^{st}$ cycle, possibly due to the semiconducting nature of the PCM deposit, resulting in a decreased current. The onset potential for GST deposition was shifted positively for the second and third cycles. The difference can be explained by the initial nucleation of the deposits during cycle 1 requiring a larger overpotential. At anodic potentials, the

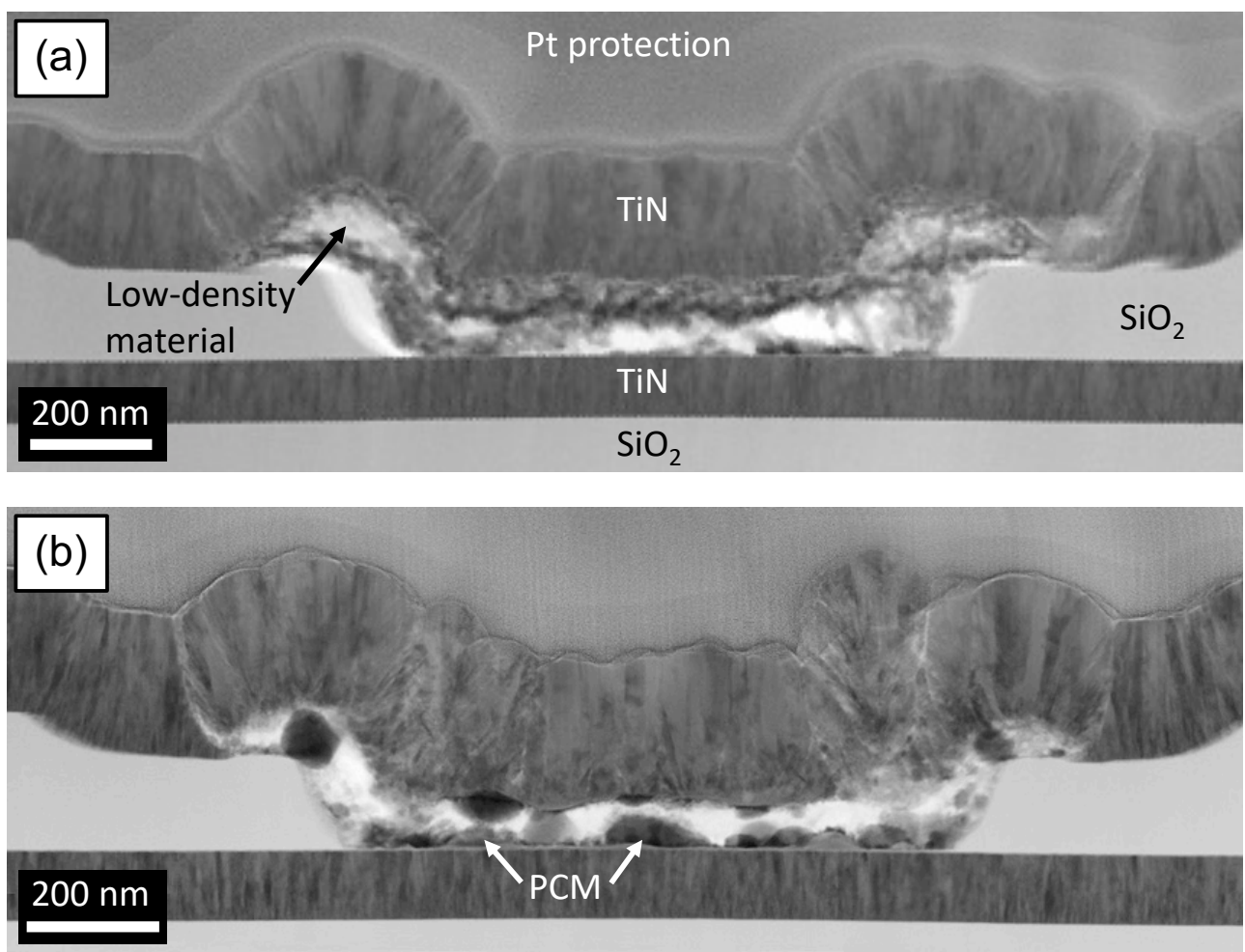

Figure 3 (a) A cross section bright field TEM image taken for one device from a 10×1 array showing the as-deposited phase change material with TiN top and bottom electrodes. (b) A cross section bright field TEM image taken from a device similar to the one shown in (a) but after it had been cycled through many phase changes.

stripping of the deposits becomes more apparent for cycles 2 and 3 compared to cycle 1 as demonstrated by their relatively higher current. Electrodepositing the PCM in the crossbar arrays was achieved by fixing the applied potential to the working electrode at -1.75 V vs Ag/AgCl and varying the total charge passed to control the deposition thickness. To minimise background currents during the electrodeposition process, the chip was placed in a purpose made sleeve which controls the surface area exposed to the solution. This was designed to prevent material deposition on the sides or back of the chip. Chronoamperometry taken during the deposition in this system can be found in Figure S4 (c). The total charge supplied to deposit the GST in the cells of a crossbar array chip is 5 µC. Theoretically, this is calculated to deposit 934 nm of GST, see the supplementary information for a breakdown of the calculation. Therefore, for a 200 nm deposit, this indicates that the Faradaic efficiency is less than 21%.

## 2. Results and Discussions

**Material Characterisation**

The PCM deposits were imaged using atomic force microscopy (AFM), as shown in Figure 2 (c) and (d). By fixing the total deposition charge to 5 µC, the deposits were found to fill the cell. Based on images of the shape of the deposits, it is clear to note that the PCM is deposited at a higher rate near the edge of a cell compared to its centre. This results in the deposits forming a "Yorkshire pudding" like shape. The rate of growth at the edges is likely to be enhanced by surface effect of the $SiO_2$ sidewalls and the change from planar diffusion to semi-spherical diffusion once the 200 nm deep cell is filled with PCM. Surface enhancement in the deposition rate near $SiO_2$ surfaces has also been reported previously.[41,42] The AFM profile presented in Figure 2 (d) shows that the shape of the sidewalls is angled. The angled shape of the sidewalls is caused by the isotropic wet etching, which was performed subsequent to dry etching in a two-step etching process (Figure S1). The central region of the deposit shows a height that is roughly equivalent to the height of the top surface of the $SiO_2$ layer, which indicates that the charge supplied during the electrodeposition is roughly equal to that required to fill the cell to its rim with PCM.

Cross sectional imaging of the phase change memory devices was performed by cutting a slice of the device using the lamella technique. Pt and C protection layers were evaporated on the sample prior to ion milling. Slices were transferred to a TEM that is equipped with EDX spectroscope to characterise electrodeposited materials as-deposited and after phase cycling. Figure 3 (a) shows a TEM cross section image taken in a bright field mode of a fabricated device. The image shows the as-electrodeposited phase change material sandwiched between the top and bottom TiN contacts. The as-deposited material appeared to be porous and contains areas of different brightness, which may indicate the presence of low-density materials.[43] These are likely to be carbon-based and maybe a result of dichloromethane reduction at high potentials. The background CV scans on DCM has been performed by several studies previously, showing its

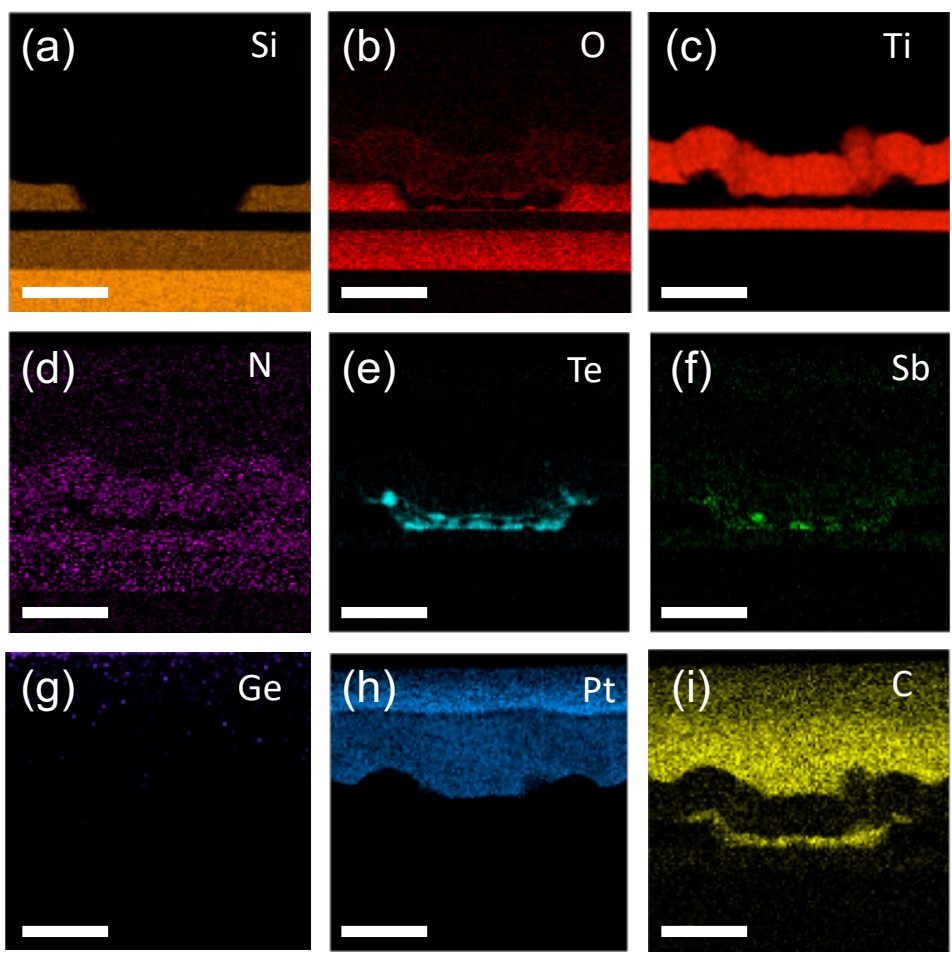

Figure 4 (a-i) Cross-sectional TEM-EDX maps of the device shown in Figure 3 (b), for the elements Si, O, Ti, N, Te, Sb, Ge, Pt and C respectively. The scale bar is 500 nm.

onset of reduction happening at around -1.7 V.[35,44] It is also clear to note here the preferential growth of the deposit along the sidewalls of the SiO$_2$ cell. The corresponding dark field images of those presented in Figure 3 can be found in the Supporting Figure S5. Figure 3 (b) shows a similar device to that from (a) but this had been phase switched around 80 times. The switching results of this particular device are presented in the next section. The PCM here is depicted by dark regions due to the heavy atomic weight of its constituents. The PCM in this device was found to have condensed into clusters of higher density materials. Electromigration of elements and void formation are common in GST based phase change memory. Xie et al. have studies this in real-time via TEM imaging.[45] In addition, significant amounts of bright, low atomic weight, material was found to be segregated from the PCM within the cell.

EDX spectroscopy was used to characterise the composition of the material. An SEM-EDX system was used to quantify the composition of the as-deposited PCM in a few microcells prior to top contact fabrication. The Ge:Sb:Te composition ratio in these deposits was found to be on average 7:8:85. Supporting Figure S6 shows an EDX spectrum of a deposit taken through top-view SEM-EDX. TEM-EDX maps were taken for the phase cycled device from Figure 3 (b) and are presented in Figure 4 (a-i). The maps show the atomic constituents of the fabricated device. Te dominance in the deposits relative to Sb and Ge is apparent in the map. The presence of C inside the cells is also clearly noted. Supporting Figure S7 shows EDX maps of the pristine device.

**Phase change memory**

The phase switching properties of electrodeposited PCMs were characterized following the fabrication of TiN top contacts in a crossbar array structure. An image of a fully fabricated 10×1 crossbar array is presented in Figure 1 (d). Phase switching the devices was performed using a Keysight B1500 that is equipped with a pulsing unit capable of supplying pulses of 10 V with 10 ns rise and fall times. The pristine devices were assumed to be electrodeposited in the amorphous state.[31] To ensure that we do not supply excessive currents that can damage the deposits or oxidize them as a result of heating, we set a system compliance current of 100 µA. We will define a device SET to be the phase switching of the electrodeposited PCM from the amorphous state to the crystalline state. In other words, a device SET will be defined as the change in resistance from a "high" to "low" state. A RESET can be defined as a switch in the phase state from crystalline to amorphous. The phase switching behavior of each device was investigated by applying voltage sweep cycles. The length of a voltage sweep is around 5s.

Figure 5 (a) shows typical SET curves from these devices, demonstrating 15 phase SET cycles. Most of the devices that have been studied as part of this work have shown SET switching between 1-2 V during the voltage sweep. After every SET cycle, the device was RESET using a 5 V pulse of width of 100 ns with rise and fall times of 100 ns. After observing several cycles from a device, endurance tests were performed by running a loop of SET-READ-RESET-READ pulses. In the endurance tests, the devices were SET using 3 V, 1 ms pulses with a rise and fall time of 1 ms and READ using 0.1 V pulses. We found that generally

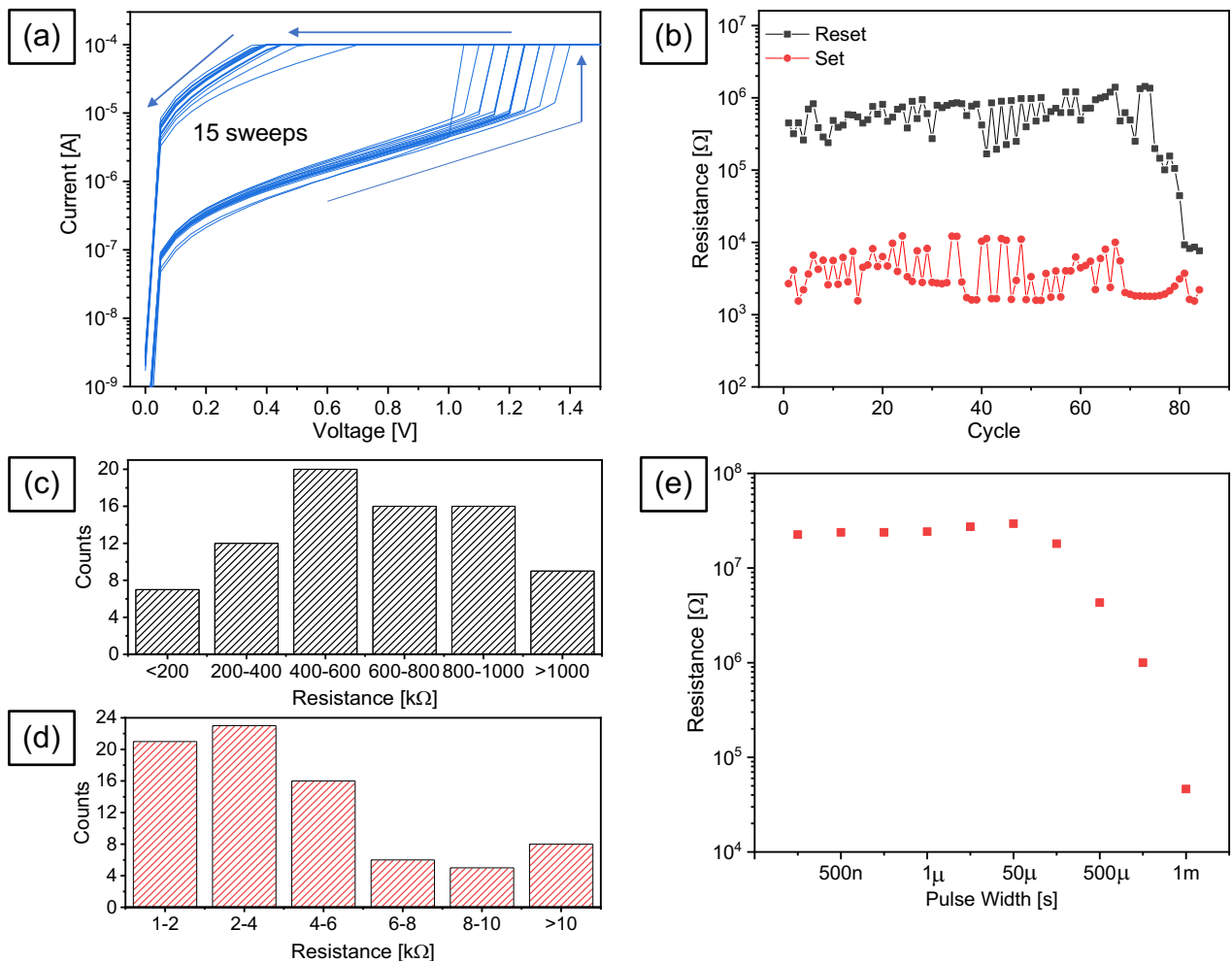

Figure 5 (a) I-V curves showing 15 voltage sweeps demonstrating phase switching of a micro device in a 10×1 array from its high resistance state (amorphous phase) to its low resistance state (crystalline phase). After every SET operation, the device was RESET using a 5V, 100 ns pulse with a 10 ns rise and fall time. (b) A graph showing the on/off resistance ratio for another device with a demonstrable endurance of around 80 cycles. Resistance distribution of the RESET (c) and SET (d) states. (e) A demonstration of the transition of a device to its low resistance state as the SET pulse width is changed showing that a pulse width greater than 50 µs is needed to induce a reduction in the device's resistance state.

the voltage required to SET a device in pulsing mode to be higher than that required to a SET the same device via a voltage sweep. Figure 5 (b) showed a device that demonstrated nearly 80 cycles before it was not possible to RESET it again. This ratio is typical for phase change devices based on GST in crossbar arrays.[46] A statistical distribution of the device resistances in SET and RESET states is presented in Figure 5 (c) and (d), respectively. Majority of the SET resistance states were in the range of 1-4 kΩ. Majority of the RESET resistance states had values in the range of 400-1000 kΩ, demonstrating over two orders of magnitude in on/off ratio. Up until the last 5 cycles, the maximum SET resistance was recorded to be 12.2 kΩ and the minimum RESET resistance state was recorded to be 100.4 kΩ. Further endurance measurements and resistance distributions from other devices are reported in Figure S8 in the Supporting Information. During our research, we have noticed a change in the composition of the deposits between multiple devices that is related to the location of the deposit on the chip. For example, deposits located at the top of the chip would be Ge richer by a few percentages than those located at the bottom of the chip. This compositional variation is possibly related to natural convection in the electrochemical cell. This can possibly be the main cause of variations in the switching cycles between multiple devices. In some cases, the endurance figures show that the devices fail in a gradual manner and remaining stuck at either of the two states. This might possibly be caused by phase segregation mediated by carbon diffusion within the deposit.

The applied pulse width was found to be critical in achieving a device SET. For example, Figure 5 (e) shows the evolution of the SET resistance as the width of the SET pulse was increased from 100 ns to 1 ms. We found that by applying pulses that are 50 µs or shorter, no significant change in the device resistance was observed. However, when the pulse width was increased beyond 50 µs, the resistance was found to decrease dramatically, reaching nearly three orders of magnitude lower values at 1 ms. We believe that there are few possible reasons which make the required pulse width to SET the devices to be relatively high. Firstly, this could be due to composition of our deposits being different from traditional GST, such as GST-225 and GST-212 compositions.[3,47] Secondly, this can also be related to the relatively large physical dimensions of our deposits, in comparison to other PCM devices from previous reports, making the thermal mass of our devices higher.[45,46,48] Achieving intermediate SET states by controlling the applied pulse width is a promising feature that makes this electrodeposited material applicable for neuromorphic computing.

This work suggests that by exploiting electrodeposition in phase change material production, true 3D memories in which the lithographic steps do not scale with the number of layers might become achievable industrially.[5] The ability to perform electrodeposition processes at room temperature limits the diffusion of materials at the metal-semiconductor interface. This allows the formation of Schottky barriers with orders of magnitude lower reversed bias current than sputtering or evaporation.[49,50] Proper implementation of a PCM-metal Schottky barrier can result in a built-in passive selector and provide the ability to efficiently scale this approach to large matrix arrays.

## 4. Conclusions

In summary, our work has shown that the non-aqueous electrodeposition technique is capable of producing GST deposits in nanofabricated crossbar structures for phase change memory applications. Crossbar bottom electrodes based on TiN were used as the working electrodes in electrodeposition. The materials were electrodeposited within 1 µm diameter cells that were fabricated by patterning a $SiO_2$ insulating layer. A non-aqueous electrolyte system was used in these electrodeposition experiments producing GST with composition of $Ge_{07}Sb_{08}Te_{85}$. Cross sectional TEM imaging of pristine and phase cycled crossbar devices have shown that the density of the deposited materials changes after many phase cycles. Electronic phase switching of several devices in 10×1 crossbar arrays have shown that an on/off ratio of around 2-3 orders of magnitude between the crystalline (SET) and amorphous (RESET) states were achievable with these materials. Endurance measurements of these devices have shown that these electrodeposited PCMs can have switching cycles reaching over 80 cycles.

## Author contributions

YJN fabricated the devices; LM and GPK performed the electrochemistry experiments and compositional analysis; AHJ performed the electrical characterization, WZ prepared the electrolytes, YH performed TEM imaging; NA performed atomic force microscopy; MA designed the electrochemical reactor; KL and NZ provided technical support in the fabrication; RH provided the original design of the chip; PNB, CHDG, GR, DCS, and RB conceived the idea and supervised the project. YJN wrote the manuscript with contributions from all authors.

## Notes

The authors declare no competing interest.

## Supporting Information

Extra details of the fabrication process, the electrochemical setup, microscopy and spectroscopy results of the deposits and electrical characterisation can be found in the supporting information document.

## Acknowledgements

The research work reported in this article was financially supported by the EPSRC programme grant Advanced Devices by ElectroPlaTing (ADEPT), grant number: EP/N035437/1.